\pdfoutput=1 
\documentclass[5p,twocolumn]{elsarticle}

\usepackage{lineno}
\usepackage{xcolor}
\usepackage{listings}
\usepackage{balance}
\definecolor{codegreen}{rgb}{0,0.6,0}
\definecolor{codegray}{rgb}{0.5,0.5,0.5}
\definecolor{codered}{rgb}{1,0,0}
\definecolor{backcolour}{rgb}{0.95,0.95,0.92}

\lstdefinestyle{mystyle}{
  commentstyle=\color{codegreen},
  keywordstyle=\color{blue},
  numberstyle=\tiny\color{codegray},
  stringstyle=\color{codered},
  basicstyle=\ttfamily\footnotesize,
  breakatwhitespace=false,         
  breaklines=true,                 
  captionpos=b,                    
  keepspaces=true,                 
  numbers=left,                    
  numbersep=4pt,                  
  showspaces=false,                
  showstringspaces=false,
  showtabs=false,
  frame=lines,
  framesep=1mm, 
  tabsize=2
}
\usepackage{titlesec}
\usepackage{amsmath}
\usepackage{flushend}

\usepackage{hyperref}

\titleformat{\paragraph}
{\itshape\normalsize}{\theparagraph .}{0.5em}{}

\modulolinenumbers[5]
\journal{Information and Software Technology}

\begin{document}

\begin{frontmatter}
\title{\textbf{DRAST - A Deep Learning and AST Based Approach for Bug Localization}}




\author[sandeepaddress]{Shubham Sangle}
\ead{cs16b026@iittp.ac.in}
\author[sandeepaddress]{Sandeep Muvva}
\ead{cs16b017@iittp.ac.in}
\author[sandeepaddress]{Sridhar Chimalakonda}
\ead{ch@iittp.ac.in}

\author[karthikeyanaddress]{Karthikeyan  Ponnalagu}
\ead{Karthikeyan.Ponnalagu@in.bosch.com, }
\author[karthikeyanaddress]{Vijendran Gopalan Venkoparao}
\ead{GopalanVijendran.Venkoparao@in.bosch.com}




\address[sandeepaddress]{Research in Intelligent Software \& Human Analytics (RISHA) Lab,\\
Department of Computer Science \& Engineering\\
Indian Institute of Technology Tirupati, India}
\address[karthikeyanaddress]{ARiSE Team, Robert Bosch Engineering and Business Solutions Ltd., Bangalore, India}




\begin{abstract}

\textbf{Context:} 
Given a bug report and source code of the project, \textit{bug localization} can help developers to focus on fixing probable buggy files rather than searching the entire source code repository. While existing research uses information retrieval (IR) and/or combination of machine learning (ML) or deep learning (DL) approaches, they focus primarily on \textit{Java} benchmark projects, and also motivate the need for multi-language bug localization approach.


\textbf{Objective:} To create a novel bug localization approach that leverages the syntactic structure of source code, bug report information and which can support multi-language projects along with a new dataset of C projects. 


\textbf{Method:} The proposed DRAST approach represents source code as code vectors by using its high-level AST and combines \textit{rVSM}, an IR technique with ML/DL models such as \textit{Random Forest} and \textit{Deep Neural Network} regressor to rank the list of buggy files. We also use features such as \textit{textual similarity} using IR techniques, \textit{lexical mismatch} using DNNs, and \textit{history} of the project using the metadata of \textit{BugC} dataset.

\textbf{Results:}
We tested DRAST on seven projects from the \textit{BugC} dataset, which consists of 2462 bug reports from 21 open-source C projects. The results show that DRAST can locate correct buggy files 90\% of the time from top 1, 5, and 10 suggested files with MAP and MRR scores of above 90\% for the randomly selected seven projects. We also tested DRAST on \textit{Tomcat} and \textit{AspectJ}, projects from benchmark dataset with better results at accuracy@1, MAP and MRR when compared with state-of-the-art. 

\textbf{Conclusions:} This paper presents a novel bug localization approach that works on C and Java projects and a bug localization C dataset along with a novel source code representation. The results for C projects using DRAST are promising and could motivate researchers/practitioners to focus on developing and creating multi-language bug localization approaches.

\end{abstract}


\end{frontmatter}


\section{Introduction}
With the advent of social coding, there is a significant rise in open-source contributions in the past decade or so. These contributions focus either on improving and enhancing the software or identifying the \textit{issues} present in the software \cite{dias2016does}. When a user faces any issues or errors while using software, the user submits a bug report to the developer team, which summarizes the scenario of the problem \cite{zhou2012should, lam2017bug}. It might contain exception thrown on the screen \cite{lukins2010bug}, stack trace, or just a textual explanation of defect \cite{wong2014boosting,lam2017bug}. There has been a rise in issue reporting, where issue tracking systems such as GitHub issue tracker allow users/contributors to report the discrepancy present in the software, allowing better software maintenance \cite{bissyande2013got}.
However, a large open-source software project typically may contain 100s/1000s of source files. Given a bug report, the task of locating the files, which are the cause for the bug, is quite tedious and effort-intensive task \cite{wong2016survey}. \textit{Bug Localization} refers to an automated process for finding the list of probable buggy files in a project related to a given bug report \cite{lam2017bug}. These automated bug localization tools are advantageous to help developers concentrate their efforts on fixing the bug from the probable list of buggy files rather than searching all files in a given repository \cite{thung2014buglocalizer}.

Researchers have investigated multiple approaches to localize bugs by proposing different ways to map the information present in a bug report to the source code \cite{lee2018bench4bl, wong2016survey}. These approaches either use Information Retrieval (IR) techniques where bug report is taken as a query and outputs a list of documents (source files) \cite{chaparro2019using} or, use Machine Learning/ Deep learning (ML/DL) techniques to reduce the ``lexical and semantic gap'' between bug reports and source code \cite{polisetty2019usefulness}. Some approaches apply the combination of IR, ML, and DL for the bug localization to bring out better results \cite{lam2017bug, xiao2017improving}. 

IR-based bug localization approaches treat the textual information present in the bug report as a query and source code as a natural text and output the ranked list of source-code files based on the textual similarity between the bug report and source files \cite{lukins2010bug, saha2013improving, youm2015bug, zhou2012should}. Along with bug reports, IR based approaches also considered class or method names \cite{saha2013improving}, version history \cite{youm2015bug}, similar bug histories \cite{zhou2012should}, structural information such as stack traces and segmentation \cite{wong2014boosting} to improve the accuracy of these approaches. ML/DL approaches usually are a combination of both IR and ML/DL techniques. Given a dataset of bug reports and source files, IR helps to establish a similarity relationship between them, and ML/DL techniques will try to map the latent features present in the bug report to source files \cite{polisetty2019usefulness, koyuncu2019d, lam2017bug, xiao2017improving, huo2019deep}. 

A common notion of evaluation for the majority of the aforementioned techniques is based on Java Datasets. \textit{Eclipse, Tomcat, SWT, JDT, AspectJ, ZXing} are the most common Java projects in bug localization approaches that are used as the benchmark \cite{polisetty2019usefulness, liang2019deep, xiao2017improving, lam2017bug, wong2014boosting, youm2015bug, wang2014version, saha2013improving, zhou2012should}. However, Shah et al. \cite{saha2014effectiveness} created a C dataset that consists of five projects to test the effectiveness of Java-based IR bug localization approaches on C projects and found that these techniques are as effective for C projects as it is for Java projects. However, we believe current approaches are not sufficient since it remains unexplored for other programming languages such as C or Python. Nevertheless, there is a need to expand the domain of bug localization approaches, which can handle other programming languages as well. In order to achieve this task, we believe new techniques should focus on the representation of the source code of different programming languages. In this paper, we are focusing on C projects available in GitHub, as C is one of the top 10 languages used in the GitHub repositories\footnote{\url{https://octoverse.github.com/}}. To the best of our knowledge, we are not aware of a bug localization approach that is developed for \textbf{C} projects that leverages both IR and ML/DL techinques.

Another important aspect we have to consider is the way source code is represented in existing bug localization approaches. In IR-based approaches \cite{youm2015bug, saha2013improving} and in some ML/DL based approaches \cite{lam2017bug, huo2019deep}, source code is considered as natural text or set of tokens as an input. Even though these approaches, especially ML/DL models, can extract lexical and syntactic information from bug reports and source code files, they are still not able to utilize the structural and semantic information present in the source code \cite{liang2019deep}. However, recent works started leveraging the structural and semantic aspects of source code in the form of abstract syntax tree (ASTs) \cite{liang2019deep, zhang2019novel}. Zhang et al. proposed ASTNN, an AST based source code representation, where source code written in C and Java is first converted into AST. It splits large AST into small statement trees and encodes these small trees into vector capturing its ``lexical and syntactic knowledge'' \cite{zhang2019novel}. CAST (customized AST) \cite{liang2019deep}, a state-of-the-art bug localization approach, where source code written in Java is first is converted to AST and then prune redundant ``syntactic entities'' before encoding them into vectors. The reason for not taking full ASTs into the model is due to its large size and requires a more extended training period \cite{liang2019deep}. One of the limitations in ASTNN is the use of pycparser\footnote{\url{https://pypi.org/project/pycparser/}}, a parser which generates ASTs for C and Java. However, due to its limited capabilities such as lots of manual pre-processing has to be done if one has to parse the whole repository, and it can handle small code snippets in a better way. The source code representation in CAST is not utilizing the source code's structure to the full extent, and it deals only with Java projects, and considers only a subset of AST. Therefore, we see a need for a novel source code representation, which takes advantage of both ASTNN and CAST approaches, and helps provide better results in the bug localization task for C and Java projects.

In this regard, we propose an approach called \textbf{DRAST} for bug localization for \textbf{C} and \textbf{Java} projects using AST representation for source code and combining Information Retrieval techniques with Machine Learning and Deep Learning models. We developed a model, \textit{\textbf{src2vec}}, to generate code vectors from source code files such that each vector represents the block of the source code by using high-level AST representation of code. \textit{src2vec} tries to reduce this limitation present in ASTNN and CAST, by representing AST into an XML format which covers each aspect of the source code. We plan to use this representation of source code in DRAST to achieve better accuracy.

DRAST computes six feature scores in which three features consider the textual similarity between source code and bug report, while the other two features consider the history of bug reports by applying information retrieval techniques. The sixth feature focuses on the problem of lexical mismatch between source code and bug report; it computes the relevancy score between a bug report and source code using Deep Neural Networks. These features are adopted from the paper on DNNLOC \cite{lam2017bug}. Finally, DRAST applies Random Forest and Deep Neural Network regressor models (the reason for selecting these models on these six features and generates a model that predicts the probability of the file being buggy for a particular bug report. It outputs the top k files, and among them, few files can be the bug source, as mentioned in the bug report. The DRAST approach is detailed in in Section \ref{section7}).  
\begin{figure}
    \centering
    \includegraphics[width=\linewidth]{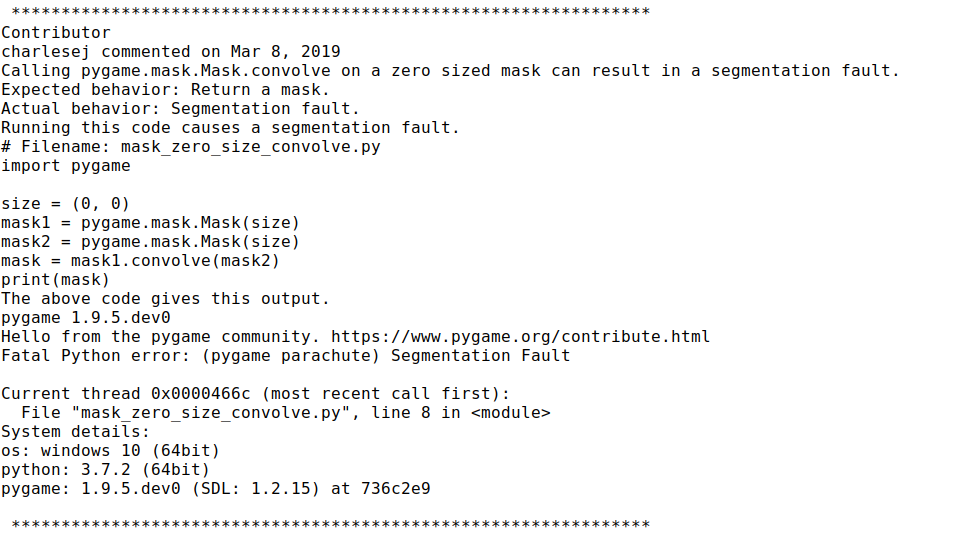}
    \caption{Sample Bug Report}
    \label{fig:bugReport}
\end{figure}

\begin{figure}
    \centering
    \includegraphics[width=\linewidth]{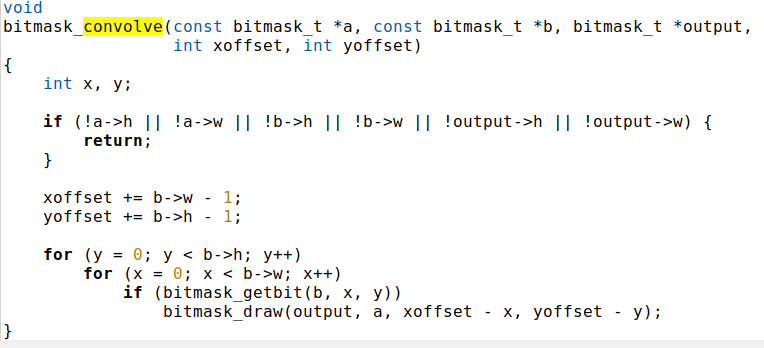}
    \caption{Buggy Function Corresponding to Bug Report in Figure \ref{fig:bugReport}}
    \label{fig:buggyFunction}
\end{figure}

\begin{figure*}
    \centering
    \includegraphics[scale = 0.57]{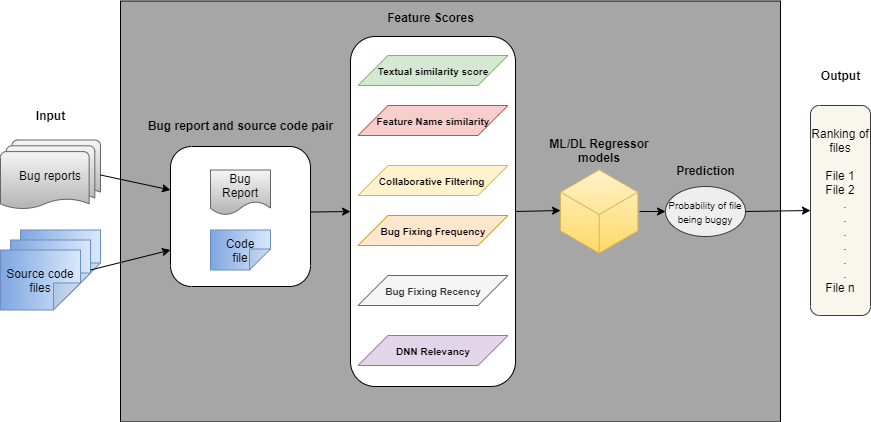}
    \caption{DRAST Pipeline}
    \label{fig:bugpipeline}
\end{figure*}

DRAST is designed to support bug localization for both C and Java-based projects. To test our approach, we have created a dataset named \textit{BugC}, which consists of 21 open-source projects written in C as a primary language. DRAST is tested on seven C projects from \textit{BugC}. DRAST is tested on seven C projects from \textit{BugC}, and it has yielded results of over 90\% for the evaluation metrics such as accuracy@1, accuracy@5, accuracy@10, see in Table \ref{tab:result1}.  We also evaluated our approach on the \textit{Tomcat} and \textit{AspectJ} project from the Java benchmark dataset, and we observed that our results are better than the other state-of-the-art approaches in terms of \textit{accuracy@1}, \textit{MAP}, \textit{MRR} (Table \ref{tab:benchmark}).

With this background, the main contributions of the paper are as follows: 
\begin{itemize}
    \item \textbf{\textit{DRAST}}: A novel approach for bug localization that supports C and Java language-based open source projects.
    \item \textbf{\textit{src2vec}}: A novel source code representation which generates code vectors of the block of the source-code file by leveraging its high-level AST representation of code.
    \item \textbf{\textit{BugC \footnote{\url{https://doi.org/10.5281/zenodo.4153560}}}}: A bug localization dataset for C projects.
    \item Evaluation of DRAST on BugC dataset and benchmark dataset
\end{itemize}

\section{Background}
\subsection{Bug Report}

Figure \ref{fig:bugReport} and \ref{fig:buggyFunction} show an example of a bug report and its related buggy file, respectively. Figure \ref{fig:buggyFunction} shows a snapshot of a code snippet, which is the source of the bug mentioned bug report. The bug report is taken from a C project - \textit{Pygame}\footnote{\url{https://github.com/pygame/pygame/issues/890}}, and it describes a segmentation fault error that occurs while invoking the \textit{convolve} function on a zero-sized mask, and the error is generated during the execution of ``\textit{mask\_zero\_size\_convolve.py}''. Hence the task of a bug localization approach is to find out the origin of the bug. The code snippet shown in Figure \ref{fig:buggyFunction} is the associated \textit{convolve} function definition, which is written in the ``\textit{bitmask.c}'' file having all associated variables and methods of ``\textit{mask}''. This function's arguments is the cause for the bug reported in Figure \ref{fig:bugReport}, due to the access of internal variable ``\textit{h}'' of \textit{NULL bitmask} ``\textit{b}'', which eventually led to a segmentation fault. This bug can be corrected by first checking whether the argument \textit{bitmask} is \textit{NULL}. In the case mentioned above, the bug was tracked by matching textual similarity. However, this might not be the case for every bug report where the information present is limited. Hence, there is a need for the bug localization approaches to map the text present in the bug report with the source code by leveraging the syntactic structure presented in a programming language.  

\subsection{Deep Neural Network} \label{dnnreg}
A deep neural network (DNN) is a multilayered artificial neural network (ANN) having one or more hidden layers between the input and output layers \cite{lam2017bug}. DNNs are feed-forward networks in which data flows without looping back from the input layer to the output layer. DNN generates a map of virtual neurons and assigns random numerical values, or ``weights,'' to the interconnections. Weights and Inputs are multiplied, which returns the output between 0 and 1. Adjustment in the weights will be repeated to perform the computation if the model does not perform accurately. This operation is continuously repeated until DNN finds relevant weights for neurons that provide better model accuracy \cite{lam2017bug}.

\subsection{Random Forest Models} \label{rfreg}
Random forest model is an ensemble model that does not require any hyper-parameters tuning and has a better performance due to its property of randomness for feature selection \cite{xiao2018bug}. It builds a forest from an ensemble of decision trees. Random forest models mostly use the method of bagging to train on the training dataset \cite{huo2016learning}.
Random forest models can be used for classifications as well as regression problems. While trees grow, random forest adds randomness to the model. While splitting a node, rather than looking for the most important feature, it looks for the best feature among a random subset of features; this results in a broad diversity, which generally leads to a better model \cite{xiao2018bug}.

\section{DRAST Model}
Figure \ref{fig:bugpipeline} shows the pipeline diagram of DRAST. A pair of bug report \textit{b} and file \textit{f} from the BugC dataset is considered as input which lists the probability of that file being buggy. For all pairs of bug report \textit{b} and file \textit{f} in training data (see Section \ref{datasetModel}), DRAST calculates a total of six features. It determines the \textit{Textual Similarity Score}, \textit{Collaborative Filtering Score}, and \textit{Feature Name Similarity Score}, which deal with the textual similarity between source code and bug reports. \textit{Bug Fixing Frequency} and \textit{Bug Fixing Recency} Score utilize the historical data of bug fixing. While in the \textit{DNN Relevancy Score}, the DNN model tries to find the relationships among bug report and source code to reduce the lexical mismatch due to their different nature. The bug report is converted using a \textit{tf-idf} vectorizer, while the conversion of source code into vectors takes place using the new approach called \textit{src2vec} (see Section \ref{codeBlocks}). The DRAST model is trained by considering these feature scores and their expected output scores. We consider our approach as a regression problem; therefore, we tried various regressor models such as xgboost, logistic regression, SVM, and so on, and chose the models that gave the best accuracy for our task. Based on our experiments (Section \ref{section7}), we selected Random Forest Regressor and DNN Regressor for DRAST because these models gave the best results for our task by successfully forming the relationship between bug reports and source code files. All the experiments and their results, are explained in Section \ref{section7}.

\begin{figure}
    \centering
    \includegraphics[width=\linewidth]{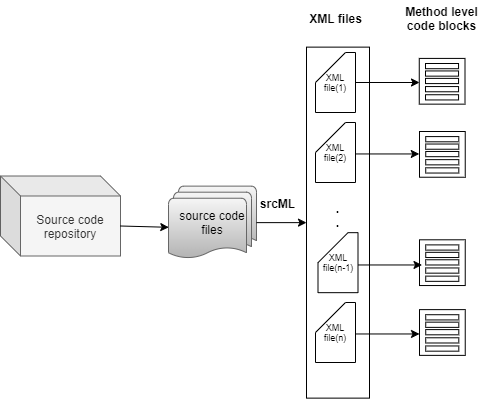}
    \caption{Code Blocks Generation}
    \label{fig:code_bloks_generation}
\end{figure}

\section{src2vec: Code Blocks Generation}\label{codeBlocks}
In this section, we will discuss the process of extracting code characteristics from the source code using its high-level AST structure. Section \ref{sec:DNN_Rel_Score} describes the process of conversion of code characteristic of source code into vectors. These code vectors will be used to calculate the features of the DRAST model (Section \ref{featureExtraction}).

\begin{lstlisting}[language=C, caption= Sample C file, label=lst:sample_c_file]
int main() { 
    FILE *fp; 
    int count = 0;  
    char filename[MAX_FILE_NAME]; 
    char c;  
  
    printf("Enter file name: "); 
    scanf("%s", filename); 
  
    fp = fopen(filename, "r"); 
  
    if (fp == NULL) { 
        printf("Could not open file %s", filename); 
        return 0; 
    } 
    for (c = getc(fp); c != EOF; c = getc(fp)) 
        if (c == '\n') 
            count = count + 1; 
  
    fclose(fp); 
    printf("The file %s has %d lines\n ", filename, count); 
  
    return 0; 
} 
\end{lstlisting}

\begin{lstlisting}[language=C, caption= Code blocks from C file, label=lst:c_code_blocks]
{"test_astnn.c": [["Decl", "FILE *", "fp", "Decl", "int", "count", "0", "Decl", "ArryDecl", "char", "filename", "MAX_FILE_NAME", "Decl", "char", "c", "FuncCall", "printf", "ExprList", "\"Enter file name: \"", "FuncCall", "scanf", "ExprList", "\" %s\"", "filename", "fp", "=", "FuncCall", "fopen", "ExprList", "filename", "\"r\"", "If", "fp", "==", "NULL", "FuncCall", "printf", "ExprList", "\"Could not open file %s\"", "filename", "return", "0", "For", "c", "=", "FuncCall", "getc", "ExprList", "fp", "c", "!=", "EOF", "c", "=", "FuncCall", "getc", "ExprList", "fp", "If", "c", "==", "'\\n'", "count", "=", "count", "+", "1", "FuncCall", "fclose", "ExprList", "fp", "FuncCall", "printf", "ExprList", "\"The file %s has %d lines\\n \"", "filename", "count", "return", "0"]]}
\end{lstlisting}

\subsection{srcML} \label{srcml}
\textbf{srcML} is a commonly used tool to represent C/C++/Java source code in XML format \cite{collard2013srcml}. The srcML format is a source code XML representation, where the markup tags identify elements of the language's abstract syntax. The srcML software is a command-line application for converting source code to srcML and vice-versa. The srcML infrastructure enables the user to further analyse, explore and modify source code. Liang et al. \cite{liang2019deep} stated that we should not use entire ASTs generated from source code due it's a large size and high processing time. Hence, we used srcML markups since they are selective at high AST level (i.e., no sub-expression) and still do not lose any information from the source code \cite{collard2013srcml}. Therefore, we used srcML to generate program vectors from source code. 

\begin{lstlisting}[language=XML, caption= XML format of C file, label=lst:c_xml_file]
  <?xml version="1.0" encoding="UTF-8" standalone="true"?> 
- <unit language="C" filename="test_astnn.c" revision="1.0.0" xmlns="littp://www.srcML.orgisrcML/src"> 
    - <function> 
     - <type> 
        <name>int</name> 
      </type> 
      <name>main</name> 
      <parameter_list>()</parameter_list> 
     - <block> 
        { 
        - <block_content> 
            + <decl_stmt> 
            + <decl_stmt> 
            + <decl_stmt> 
            - <decl_stmt> 
                - <decl> 
                    - <type> 
                        <name>char</name> 
                      </type> 
                      <name>c</name> 
                  </decl> 
                  ;
              </decl_stmt> 
            + <expr_stmt> 
            + <expr_stmt> 
            + <expr_stmt> 
            + <if_stmt> 
            + <for> 
            + <expr_stmt> 
            + <expr_stmt> 
            + <return> 
          </block_content> 
        } 
     </block> 
    </function> 
</unit> 
\end{lstlisting}

\subsection{Conversion of XML Files to Code Blocks}
Following the documentation\footnote{\url{https://www.srcml.org/}}, we have written a python script that creates XML files using the srcML command-line tool in their respective folder path and parse the created XML file to convert the source code file into code blocks/vectors. The script is divided into two files. The first file has all the function definitions to convert a particular source code file to XML file representation and convert it into code blocks/vectors (Figure \ref{fig:code_bloks_generation}), and extract code characteristics from it. The second file traverses through the whole project folder and creates a folder with the name projectName\_XML and stores all the generated XML files in their respective paths as the original project folder. It also stores the extracted code blocks/vectors and code characteristics in different formats. Figure \ref{fig:code_bloks_generation} shows the generation of code blocks for a project repository. 

\begin{lstlisting}[language=Java, caption= Sample Java file, label=lst:sample_java_file]
import java.util.Scanner;
public class AddTwoNumbers2 {

    public static void main(String[] args) {
        
        int num1, num2, sum;
        Scanner sc = new Scanner(System.in);
        System.out.println("Enter First Number: ");
        num1 = sc.nextInt();
        
        System.out.println("Enter Second Number: ");
        num2 = sc.nextInt();
        
        sc.close();
        sum = num1 + num2;
        System.out.println("Sum of these numbers: "+sum);
    }
}
\end{lstlisting}

\begin{lstlisting}[language=Java, caption= Code blocks from Java file, label=lst:java_code_blocks]
{"java folder\\test.java": [
["import", "java", "util", "Scanner"],
["class", "public", "AddTwoNumbers2", "block"], 
["function", "type", "public", "static", "void", "main", "parameters", "type", "String", "args", "block", "declaration", "type", "int", "num1", "type", "num2", "type", "sum", "declaration", "type", "Scanner", "sc", "expression", "call", "System", "out", "println", "arguments", "expression", "expression", "num1", "call", "sc", "nextInt", "arguments", "expression", "call", "System", "out", "println", "arguments", "expression", "expression", "num2", "call", "sc", "nextInt", "arguments", "expression", "call", "sc", "close", "arguments", "expression", "sum", "num1", "num2", "expression", "call", "System", "out", "println", "arguments", "expression", "sum"]
]}
\end{lstlisting}

  
  
  
  
  

For C and Java projects we considered only those files that end with the extension \textit{.c} and \textit{.java} respectively. In the C project, each C file is divided into code blocks/vectors based on the function definition, whereas in a Java project, each java file is divided into code blocks/vectors based on function and class definition. Listing \ref{lst:sample_c_file}, \ref{lst:c_code_blocks}, and \ref{lst:c_xml_file} shows a sample C file, code blocks extracted and file's XML format representation using srcML  respectively. Similarly, Listing \ref{lst:sample_java_file},\ref{lst:java_code_blocks}, and \ref{lst:java_xml_file} shows a sample for a Java file.


\subsection{Extracting Code Characteristics from Source Code File} \label{codeFeatures}
During the conversion of the source code file into code blocks/vectors, we also extracted the characteristics from the source code that are useful in calculating the DNN relevancy score, which is one of the features for our final Bug Localization model. In C projects, we extracted the characteristics, namely \texttt{function names}, \textit{identifiers}, \textit{macros}, \textit{unions}, \textit{typedef}, \textit{struct}, \textit{cpp} from the C file. Likewise, from Java projects \textit{function names}, \textit{class names}, \textit{identifiers} are extracted. Listings \ref{lst:c_code_features}, \ref{lst:java_code_features} represent the code features extracted from the sample C file and Java file respectively. 
\begin{lstlisting}[language=XML, caption= XML format of Java file, label=lst:java_xml_file]
 <?xml version="1.0" encoding="UTF-8" standalone="true"?>
- <unit language="Java" filename="java folder\test.java" revision="1.0.0" xmlns="http://www.srcML.org/srcmL/src">
    - <import>
        import 
        + <name>
        ;
      </import>
    - <class> 
        <specifier> public</specifier> 
        class 
        <name>AddTwomumbers2</name> 
        - <block> 
            { 
          - <function> 
             + <type> 
             <name>main</name> 
             + <parameter_list> 
             - <block> 
                 {
                - <block_content> 
                    + <decl_stmt> 
                    + <decl_stmt> 
                    + <expr_stmt> 
                    + <expr_stmt> 
                    + <expr_stmt> 
                    + <expr_stmt> 
                    + <expr_stmt> 
                    + <expr_stmt> 
                    + <expr_stmt> 
                  </block_content> 
                  }
                </block>
            </function> 
            } 
        </block> 
    </class> 
</unit> 


\end{lstlisting}
\begin{lstlisting}[language=C, caption= Code features extracted from C file, label=lst:c_code_features]
{"test_astnn.c": {"function": ["main", "printf", "scanf", "fopen", "printf", "getc", "getc", "fclose", "printf"], "struct": [], "cpp": [], "macro": [], "identifier": ["fp", "count", "filename", "MAX_FILE_NAME", "c", "filename", "fp", "filename", "fp", "NULL", "filename", "c", "fp", "c", "EOF", "c", "fp", "c", "count", "count", "fp", "filename", "count"], "typedef": [], "union": []}}
\end{lstlisting}




\begin{lstlisting}[language=Java, caption= Code features extracted from Java file, label=lst:java_code_features]
{"java folder\\test.java": ["java", "util", "Scanner", "AddTwoNumbers2", "void", "main", "String", "args", "int", "num1", "num2", "sum", "Scanner", "sc", "System", "out", "println", "num1", "sc", "nextInt", "System", "out", "println", "num2", "sc", "nextInt", "sc", "close", "sum", "num1", "num2", "System", "out", "println", "sum"]}
\end{lstlisting}

\section{Feature Generation} \label{featureGen}
This section explains about the input preprocessing and the feature extraction process in the DRAST approach. 

\subsection{Input Pre-processing} \label{preprocessing}
Bug report may contain many words or elements that do not contribute towards bug localization, so to remove those elements and extract maximum possible information from a given bug report, we perform various pre-processing tasks:

\begin{itemize}
    \item \textbf{Tokenization}: The process of splitting up a sequence of strings or text into words or symbols, commonly known as tokens, is called tokenization. We are tokenizing the bug report by considering space between words as a separator.
    \item \textbf{Removing Stop Words}\label{stopWords}: Stop words are a group of words present in every language, which do not contribute towards it is meaning but are very helpful for humans for understanding language. For many applications such as classification or clustering, removing stop words helps increase accuracy and performance because they do not provide much unique information as they occur in abundance. For example, words like “a,” “is,” “the” are quite common in natural language but not in the programming language. Hence, we perform this step on the bug report to remove frequently used useless words.
    \item \textbf{Stemming Words}\label{stemWords}: Stemming is a text normalization technique used for reducing a word to its stem, which is a simpler form of the word. The aim of stemming is to reduce a word from inflectional or derivationally form to a base form.
    \item \textbf{Splitting Camel Case Words}\label{camelCase}: Camel case is a naming convention for declaring identifiers, functions, classes, and it is used in many programming languages, in which every word in a compound word is capitalized except for the first word. In this process, these words are split into independent words, and the original word is also stored along with them.
\end{itemize}
\begin{figure*}
    \centering
    \includegraphics[width=\linewidth]{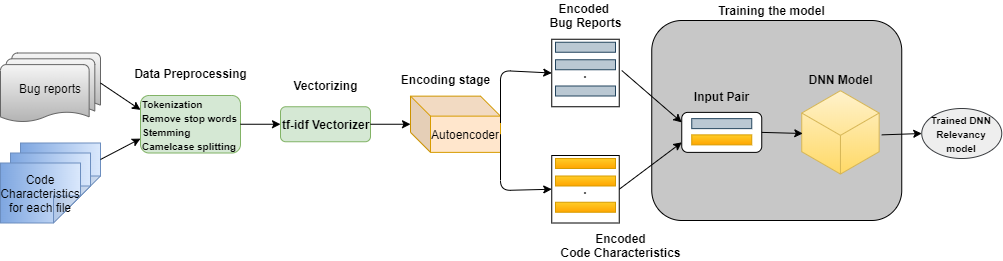}
    \caption{DNN Relevancy Model}
    \label{fig:dnn_relevancy_model}
\end{figure*}
\textbf{Pre-processing Program Vectors}\label{preprocessing code features}: Program vectors which are generated from each source code file by performing operations discussed in Section \ref{codeBlocks} are already in tokenized form. Therefore, other pre-processing steps mentioned above are also performed on these program vectors.

\begin{figure}
    \centering
    \includegraphics[width=\linewidth]{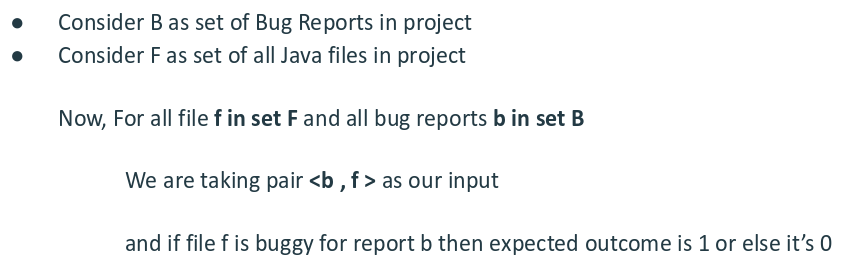}
    \caption{Methodology For Creating Pairs}
    \label{fig:input}
\end{figure}

\subsection{Feature Extraction} \label{featureExtraction}
For creating a dataset for training and testing our model, we create positive as well as negative pairs between a bug report and source code file, as shown in Figure \ref{fig:input}. Positive pairs are created by pairing bug reports with their corresponding buggy files, while negative pairs are created by pairing bug reports with the rest of the source code files.
\par

For each pair of bug report \textit{b} and file \textit{f} following feature scores are computed. These features are adopted from DNNLOC \cite{lam2017bug}, which are as follows:

\subsubsection{Textual Similarity Score} \label{textSimilarity}
Textual similarity score is calculated for every code block in file \textit{f} and bug report \textit{b} to find the semantic similarity between them and used the \textit{revised Vector Space Model} (rVSM) model for this purpose \cite{zhou2012should}. rVSM ranks larger files higher because they are more prone to be buggy \cite{ostrand2005predicting,fenton2000quantitative}.

For each source file \textit{f}, there are multiple program vectors. Textual similarity score is computed between each program vector in file \textit{f} and bug report \textit{b}. Maximum similarity score among them is taken as final textual similarity score for pair of file \textit{f} and bug report \textit{b}.

\subsubsection{Collaborative Filtering Score} \label{colab_filter}
Collaborative filtering score is calculated if similar bug reports are fixed in file \textit{f} before bug report \textit{b}, then there is higher chance that file \textit{f} is again buggy for bug report \textit{b}. The score shows the textual similarity score between bug report \textit{b} and earlier bug reports, which was fixed in file \textit{f}.

\subsubsection{Feature Name Similarity}\label{featureNameSim}
Bug reports often contain error messages mentioning the name of class, function, API, and so on. These crucial terms might be used only once or twice in the entire bug report and source code. When normal textual similarity score is computed, there is a chance that the low score is calculated for file \textit{f} in which these classes or functions names are present, and still \textit{f} is buggy for bug report \textit{b}. To overcome this issue, we compute the \textit{feature name similarity} score. To calculate the score, code characteristics (Section \ref{codeFeatures}) are extracted for file \textit{f}. Textual similarity score is calculated between these code characteristic for file \textit{f} and bug report \textit{b}.

\subsubsection{Bug Fixing Recency} \label{bugFixRec}
Kim et al. suggest that files that were buggy more recently have higher chances to be buggy again than the other files \cite{kim2007predicting}. 

If bug report \textit{b'} is most recent bug which is fixed in the file \textit{f} before bug report \textit{b} is recorded then, bug fixing recency is reciprocal of difference of \textit{months} between timestamp of bug report \textit{b} and bug report \textit{b'}. This will give a higher score to file, which is fixed more recently compare to file, which is not fixed for a large amount of time.

\subsubsection{Bug Fixing Frequency}
Files which are fixed multiple time in the past, are prone to be buggy again in future. For pair of bug report \textit{b} and file \textit{f}. Bug fixing frequency is the number of time file \textit{f} was detected as buggy before bug \textit{b} is reported. \par

\subsubsection{DNN Relevancy Score}
\label{sec:DNN_Rel_Score}
DNN relevancy score is calculated with the expectation that DNN can recognize the relationships among the features of two different natures, one extracted from bug reports and the other from source code. Figure \ref{fig:dnn_relevancy_model} shows an overview of how the model for DNN relevancy is trained and saved. We base this core from DNNLoc \cite{lam2017bug}, and the below sections which will walk through the process involved in calculating the DNN relevancy score. 
\begin{enumerate}
    \item \textbf{Pre-processing}: Pre-processed bug reports and the extracted code characteristics of the source code (Section \ref{codeFeatures}) are taken as input and again undergoes the pre-processing steps as mentioned in Section \ref{preprocessing code features}.
    \item \textbf{Conversion into Vectors}: Pre-processed bug reports and code features are converted into \textit{tf-idf} vectors with a shared vocabulary. 
    \item \textbf{Creating Positive and Negative Pairs}: For a given bug report \textit{b}, if a file \textit{f} is buggy, then the pair of bug report vector of \textit{b} and code vector of file \textit{f} is called a positive pair; otherwise, it is called a negative pair. We observe that number of positive pairs is far less than the negative pairs because of a large number of files in the project, and only 2-5 files are buggy for each bug report. To address this imbalance issue, before the autoencoding stage, we make the number of negative pairs equal to positive pairs by sorting them according to the textual similarity score and include the top negative pairs.
    \item \textbf{Autoencoding}: As the dimensions of these vectors are high, theyc require high computation power and time. Reducing the dimensions might also help in getting rid of redundant information and can make our model scalable. To reduce the dimensions of the feature vectors, we used an autoencoder with two steps encoding and decoding. The encoding stage reduces the dimensions of the input vectors, whereas the decoding stage ensures that the reduced space does not lose any useful information. In our model, we are reducing the dimensions to 75\% of the original dimensions. This technique used by DNNLoc \cite{lam2017bug} as well. The positive and negative pairs created are encoded using this autoencoder. 
    
    \item \textbf{Feeding to DNN Relevancy Estimator}: The encoded pairs are divided into training and testing data and fed to the DNN model to calculate the DNN relevancy score. The DNN model has 3 layers, namely input, hidden, and output. We used \textit{tanh} and \textit{ReLU} activation function at the hidden layer and \textit{sigmoid} activation function at the output layer. The number of nodes at the output layer is 1, whereas the number of nodes at the hidden layer can be varied. The output of this model is our final DNN relevancy score, which ranges from 0 to 1. 
        \begin{align*}
     Sigmoid\_Activation\_Function(z) &= \frac{1}{1+\exp{-z}}  \\ 
     tanh\_Activation\_Function(z) &= \tanh z \\
     \begin{split}
     Rectifier\_Linear\_Unit(z) &= z^{+} \\
                            &= max(0,z)    
     \end{split}
    \end{align*}
    \item \textbf{Saving the Models}: For later use, vectorizer, autoencoder, and DNN relevancy estimator mentioned above are saved in the usable formats. 

\end{enumerate}

\section{Dataset}\label{dataset}

\subsection{Objectives}
Current bug localization techniques are evaluated mostly on Java projects. In 2018, Lee et al. \cite{lee2018bench4bl} created Bench4bl, which consists of 51 java projects having 10,017 bug reports. This is a reproducibility study to test the performance of IR based bug localization techniques on a larger number of test projects. However, earlier bug localization techniques/tools are not tested for programming languages other than Java. This examination motivated us to construct a dataset on a different programming language, i.e., C. This dataset could be used as a benchmark dataset for C for bug localization purposes in the future along with comparative studies with the current Java benchmark dataset.

\begin{figure}
    \centering
    \includegraphics[width=\linewidth]{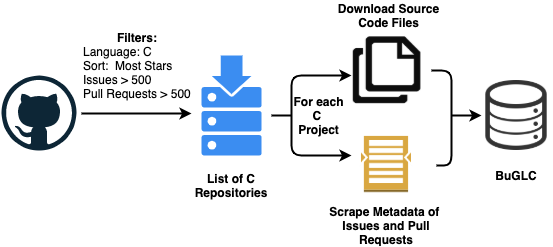}
    \caption{Methodology for Curating the Dataset}
    \label{fig:methodology}
\end{figure}

\subsection{Methodology for Choosing Projects} \label{Criteria for projects}
GitHub contains an extensive source of software artifacts such as issues and pull-requests presented in many open source projects. Features such as GitHub Stars reflect the developers' interests that are either watching or contributing to the project development, thus acting as a popularity index for GitHub projects \cite{borges2018s}. Projects are selected based on the primary programming language—GitHub's linguistic feature help to find the projects written in C (see Figure \ref{fig:methodology}). The criteria used to select a C project are: it should consist of at least 500 closed issues and pull requests. The intuition behind this criteria is that at least 100 issues could be correlated to the pull-requests (pull-request which solve the issue). This correlation provides vital information about the issues and how they are fixed.

\subsection{Methodology for filtering the bugs} \label{Criteria for Bugs}
In a general scenario, for the bug localization approach, there should be a bug report that mentions the issue and the record of files affected/changed due to this issue. However, the GitHub issue tracker does not contain metadata about how the issues have been resolved. The metadata, i.e., changes in files that rectify the issue, is vital for bug localization, and pull-requests contain this information. Still, a pull-request is not restricted to resolving issues but can also provide enhancement or new features to the code base. Hence, pull-requests cannot be considered as bug fixes by default. Although labels exist for a pull-request such as a bug, feature, fixes, enhancement, and so on, these labels cannot be seen in the majority of the pull-requests, making the identification of the pull-request type difficult. Therefore, identifying the pull-requests that resolve the issues will fetch the required metadata related to how issues have been resolved. This identified correlation between pull-request and issues is to be included in the dataset.

Issues can be correlated with pull requests based on keywords such as \textit{fixes}, \textit{resolves}, and so on, followed by \textit{\#issue ID} which can be found in the description of a pull request. If the collaborator merges the pull-request in the code base, then the issue (mentioned by issue ID) is considered to be resolved by this pull-request. The metadata from this pull request, i.e., lines/ files changed, can be used to relate to the issue. 
\begin{figure}
    \centering
    \includegraphics[width=\linewidth]{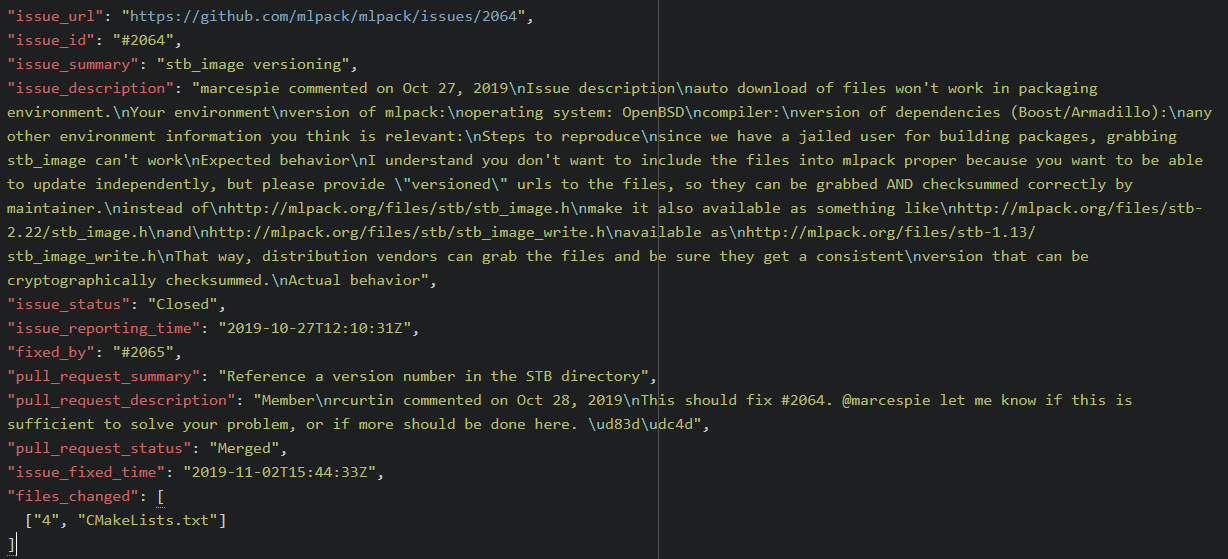}
    \caption{Metadata of a Sample Bug}
    \label{fig:sample bug}
\end{figure}

\begin{figure}
    \centering
    \includegraphics[width=\linewidth]{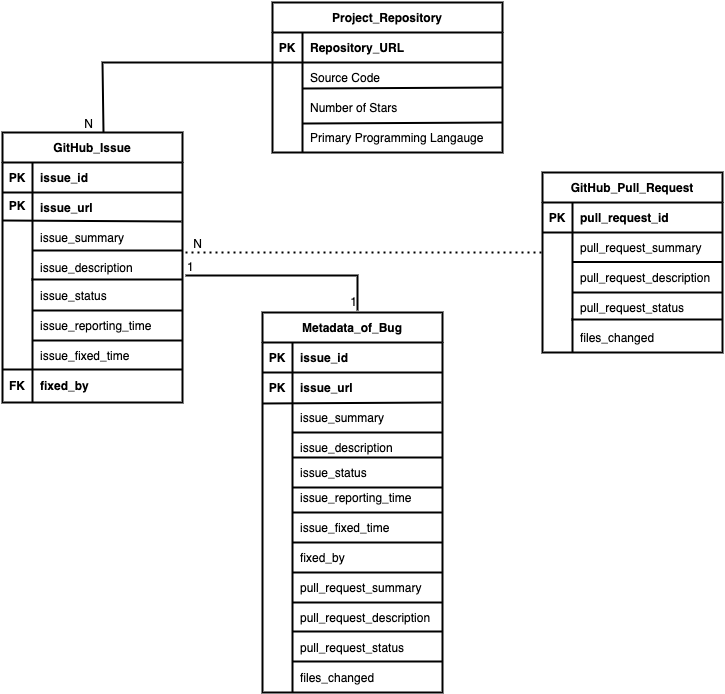}
    \caption{BugC Database Schema}
    \label{fig:database schema}
\end{figure}

\subsection{Dataset Creation}
C projects are extracted from GitHub based on a three-step selection process, presented below:

\begin{itemize}
    \item Using GitHub's linguistic feature projects with C as the primary programming language is selected.
    \item Descending sort is performed based on the number of stars.
    \item Manual selection of C projects based on selection criteria mentioned in Section \ref{Criteria for projects}
\end{itemize}

Figure \ref{fig:methodology} summarizes the methodology adopted for curating the dataset. BugC consists of metadata and source code projects of 21 C projects from GitHub. In the next step, metadata of correlated issues and pull requests are collected. We executed a python script and used selenium and chrome driver to scrape the required metadata such as \textit{issue id}, \textit{description}, and so on, and then stored it in \textbf{json} and \textbf{xlsx} formats. Figure \ref{fig:sample bug} shows a sample bug that contains metadata such as \textit{issue id}, \textit{issue summary}, \textit{issue description}, \textit{issue reporting time}, \textit{issue status}, \textit{fixed by (pull-request ID)}, \textit{pull-request description}, \textit{pull-request status}, \textit{files changed}, \textit{lines changed} in each file. For open issues, metadata related to pull requests is empty. Figure \ref{fig:database schema} summarizes the schema of BugC. The schema has four major components, i.e., \texttt{Project\_Repository} which stores metadata related to the repository, \texttt{GitHub\_Issue} stores metadata of the repository's issues, \texttt{GitHub\_Pull\_Request} stores information related to the repository's pull requests, and \texttt{Metadata\_of\_Bug} which have the information regarding correlated issues and pull requests. Table \ref{tab:final stats} summarizes stats for BugC. An important thing to note here is, in the BugC dataset, even though there are more than 36000 closed issues, we are considering only those issues as bug reports for the bug localization purpose, which have been resolved by pull requests. Because to train and test DRAST, we need bug reports which are already being resolved. The rest of the closed issues are either resolved by the core development team, in which they do not have to push the pull requests or through online discussions. Hence, in the end, we have the data of only 2462 bug reports, which are resolved via pull-requests for 21 C projects. 

\begin{table}[!htbp]
    \caption{Statistics of BugC dataset}
    \label{tab:final stats}
    \centering
    \begin{tabular}{|p{0.7\linewidth}|c|}
        \hline
        Number of Projects & 21 \\
        \hline
        Number of Closed Issues & 36617 \\
        \hline
        Average Number of Issues & 1744 \\
        \hline
        Total Number of Issues Closed by Pull Requests & 2462 \\
        \hline
    \end{tabular}
    
\end{table}

\section{Dataset creation for models} \label{datasetModel}

Consider B as a set of all bug reports \textit{b} reported till now, and F as a set of all files \textit{f} for a particular project. Here we pair all bug reports with each file in a project and compute six feature scores for each pair, as mentioned in Section \ref{featureGen}. Since supervised learning models are used, some expected output for every pair of bug report \textit{b} and file \textit{f} should be present. In a pair \textbf{\textless b,f\textgreater} if \textit{f} is a buggy file corresponding to bug report \textit{b}, then we consider predicted output for that pair as 1 or else we consider predicted output to be 0. Also, \textbf{\textless b,f\textgreater} = 1 then it is known as positive pair, otherwise \textbf{\textless b,f\textgreater} = 0 considered as negative pair.

As there can be thousands of files in large projects, training our model as well as computing all six features on all those pairs will be a performance-intensive task and can take considerable training time. We decided to set a threshold; we considered only those pairs which have a textual similarity score greater than 0.1 (this number has been arrived after empirically verifying multiple cut-offs). The decision to select a threshold lies in the datasets' nature, which is a highly imbalanced one (the number of negative pairs is more than the positive pairs). Earlier studies such as DNNLoc \cite{lam2017bug} handled the imbalanced dataset by selecting all the positive pairs, and they use textual similarity scores between bug reports and source code files to undersample negative pairs and rank them in descending order and select the top 300 files. While NP-CNN \cite{huo2016learning} uses undersampling operation on negative pairs of the training datasets to make the number of negative pairs equal to positive pairs.

 All the pairs satisfying threshold condition and along with their six computed feature scores and predicted outputs, form our dataset for training and testing the model. We sort these pairs in ascending order of timestamp so that the older bug will be ranked higher, and then we divide the dataset in k folds, where k is decided according to the size of the project. We choose to keep around 100 bug reports and their all pairs in one fold.

\section {Evaluation metrics} \label{metrics}
The following metrics are considered to evaluate DRAST's performance as these metrics are commonly used by the research community to evaluate bug localization approaches.
\begin{enumerate}
    \item \textbf{Accuracy@k}: Accuracy@k metrics indicate the probability of finding an actual buggy file for a given bug report in top k files predicted by DRAST. For each bug report, all files are sorted in descending order of final score predicted by DRAST. We considered the range for the value of k between 1 to 15. If for bug report \textit{b}, any of it is corresponding buggy file is found in top k files after sorting, then we count that as a success, or else we count it as a miss. Accuracy@k will be considered as a ratio of success to the total number of bug reports.
    \item \textbf{Mean Average Precision}: MAP (Mean Average Precision) is a mean of average precision values calculated for all predictions corresponding to each bug report. Average precision is mean of the ratio of actual buggy files in top k files predicted by our model over k. 
\begin{align*}
    AveragePrecision(b) &= \sum_{k \in K}^{} \frac{Precision(b)}{|K|} 
\end{align*}
\begin{align*}
    MAP &= \sum_{b=1}^{|B|} \frac{AveragePrecision(b)}{|B|}
\end{align*}
    \item \textbf{Mean Reciprocal Rank}: MRR (Mean Reciprocal Rank) is a mean of reciprocal of the rank of the actual buggy file in the list of files predicted by our model. This metric helps us to evaluate the overall performance of the model for all buggy files on each and every bug.
\begin{align*}
  MRR &= \frac{1}{|TestSet|} \sum_{i=1}^{|TestSet|} \frac{1}{index(i)} \\  
\end{align*} 

\end{enumerate}

\section{Experiments} \label{section7}
There are many regression models in machine learning, but to find the best possible model suitable for our task of bug localization, we performed multiple experiments with various models and recorded their performance according to the performance metrics (Section \ref{metrics}). All the experiments are performed on \textbf{\textit{Tomcat}} project, which is a project from the Java benchmark dataset and used in nearly all state of the art bug localization models for evaluation purposes \cite{zhou2012should, lam2017bug, liang2019deep, huo2016learning, huo2017enhancing}. We chose this project as it is smaller in size compared to other projects in the benchmark dataset and has a large number of bugs in the dataset to train the model. It has a total of 1056 bug reports in the dataset and a total of 2480 java source code files in total.
 
\subsection {Train-Test data} \label{trainTest}
As discussed in Section \ref{datasetModel}, k folds of data are created. Consider those folds to be fold\textsubscript{1}, fold\textsubscript{2}, ..., fold\textsubscript{k}. These folds are sorted according to their timestamp. Hence, bug reports in fold\textsubscript{i} are older than bug reports in fold\textsubscript{i+1}. The model is trained on older bug reports and test it on newer bug reports. Therefore the training is done using fold\textsubscript{i} and evaluation is done against fold\textsubscript{i+1}. 

\begin{table*}[!htbp]
     \caption{Oversampling-Undersampling Experiment Results on Tomcat Project}
    \label{tab:exp1}
    \centering
    \begin{tabular}{|c|p{0.25\linewidth}|c|c|c|c|c|}
    \hline
      \textbf{\textit{ Model Name}}  &  \textbf{\textit{Oversampling / Undersampling technique used}} & \textbf{\textit{accuracy@1}} & \textbf{\textit{accuracy@5}} & \textbf{\textit{accuracy@10}} & \textbf{\textit{MAP}} & \textbf{\textit{MRR}} \\ \hline
        Random Forest & No oversampling & \textbf{64.51}\% & \textbf{67.74}\% & \textbf{72.04}\% & \textbf{0.588} & \textbf{0.676} \\ \cline{2-7}
         & SMOTE oversampling &	\textbf{61.29}\%	 &	\textbf{67.74}\% &	\textbf{69.89}\%	&	\textbf{0.553} & \textbf{0.645} \\ \cline{2-7}
         &	ADASYN oversampling &	59.13\% &	65.59\% &	67.74\% &	0.535 &	0.622 \\ \cline{2-7}
         &	Kmeans SMOTE oversampling &	52.68\% &	66.66\% &	68.81\% &	0.51 &	0.584 \\ \cline{2-7}
         &	Random undersampling &	59.13\% &	67.74\% &	73.11\% &	0.557 &	0.64 \\ \cline{2-7}
         &	TOMEK links undersampling &	64.51\% &	67.34\% &	71.06\% &	0.582 &	0.669 \\ \hline
        DNN & No oversampling & \textbf{68.81\%} & \textbf{73.11\%} & \textbf{75.26\%} & \textbf{0.612} & \textbf{0.708} \\ \cline{2-7}
         &	SMOTE oversampling	& \textbf{39.78}\% &	\textbf{58.06}\% &	\textbf{66.66}\% &	\textbf{0.439}	& \textbf{0.497} \\ \cline{2-7}
         &	ADASYN oversampling &	30.10\% &	51.61\% &	56.98\% &	0.369 &	0.407 \\ \cline{2-7}
         &	Kmeans SMOTE oversampling &	34.40\% &	56.98\% &	68.81\% &	0.408 &	0.45 \\ \cline{2-7}
         & 	Random undersampling &	32.25\% &	52.68\% &	59.13\% &	0.354 &	0.414 \\ \cline{2-7}
         &	TOMEK links undersampling &	53.76\% &	65.59\% &	67.74\% &	0.506 &	0.583 \\ 
         \hline
        
    \end{tabular} \\
   
\end{table*}

\subsection {Over-Sampling and Under-Sampling} \label{over-under-exp}
While creating the dataset for Tomcat, we observed huge numbers of negative pairs compared to positive pairs. There are a total of 1823 positive pairs, whereas there are nearly 302920 negative pairs in the dataset. There is class imbalance, with the imbalance ratio as 1:166. There is a chance that the model might give high accuracy, but we might get a low precision-recall score due to this imbalance. So, we performed experiments by training our model on imbalanced data as well as synthetic data created by various over-sampling methods and under-sampling methods. 

Over-sampling and under-sampling are methods used to remove class imbalance from the dataset. In over-sampling, the minority class population is increased by generating synthetic data by various methods or randomly multiplying existing elements of the minority class. In under-sampling, the majority class population is reduced by randomly removing some of the data elements from the majority class or removing very similar types of data elements from the dataset. In the following list, we described the over-sampling and under-sampling techniques applied on training data and fed to Random Forest regressor and DNN regressor:
\begin{itemize}
    \item \textbf{Synthetic Minority Over-Sampling Technique (SMOTE)}: SMOTE  method \cite{chawla2002smote}, `` first selects a minority class instance at random and finds its k nearest minority class neighbors. The synthetic instance is then created by choosing one of the k nearest neighbors b at random and connecting a and b to form a line segment in the feature space. The synthetic instances are generated as a convex combination of the two chosen instances a and b'' \cite{he2013imbalanced}. We apply the SMOTE technique on our dataset to synthetically generate positive pairs for the dataset and remove imbalance from our training data.
    \item \textbf{Adaptive Synthetic Sampling}: He et al. \cite{he2008adasyn} developed an Adaptive Synthetic Sampling (ADASYN) model, an oversampling model to increase the population of the minority class. ADASYN is an improved version of SMOTE. After generating synthetic data, the ADASYN model adds very small values to make synthetic data more realistic and scatter data to avoid overfitting. 
    \item \textbf{Random Over-Sampling and Under-Sampling}: Random over-sampling is a technique in which data elements are chosen randomly from training data, and it's multiple copies are created. Random under-sampling is a technique in which data elements are randomly chosen and removed from training data to reduce the class imbalance.
    \item \textbf{Kmeans SMOTE Over-Sampling}: This technique is a combination of both the Kmeans clustering algorithm and SMOTE technique. With the availability of clusters, SMOTE can identify the target cluster areas where artificial data can be more effective and, therefore, able to reduce noisy samples \cite{douzas2018improving}. 
    \item \textbf{TOMEK links Under-Sampling}: In this technique, clusters are formed using an imbalanced dataset, and TOMEK links are used to remove the overlapped data points between the classes by removing majority class links \cite{kotsiantis2006handling}. This process leads to well-defined clusters, which can be useful for training the model.
\end{itemize}

Table \ref{tab:exp1} shows results of the all the experiments performed on \textit{Tomcat} project. From this table, we can observe that \textit{No Oversampling} and in the oversampling techniques \textit{SMOTE} yields promising results with our dataset and models. So, we decided to use the SMOTE method to handle the class imbalance in further experiments.

\subsection {Regression Models for Bug Localization} \label{sec: Regressor_model}
There are various machine learning models that are quite often used for regression tasks, which involved multiple parameters such as LASSO regression, Elastic Net regression, Gradient Boost, XGBoost, and LightGBM along with Random Forest and Deep Neural Network regressors. \par

\begin{table*}[!htbp]
    \caption{Various Regressor Models results on Tomcat project}
    \label{tab:exp2}
    \centering
    \begin{tabular}{|p{0.25\linewidth}|c|c|c|c|c|c|}
    \hline
        \textbf{\textit{ Oversampling technique used}}  & \textbf{\textit{Model Name}} & \textbf{\textit{accuracy@1}} & \textbf{\textit{accuracy@5}} & \textbf{\textit{accuracy@10}} & \textbf{\textit{MAP}} & \textbf{\textit{MRR}} \\ \hline
        No Oversampling  &	Random Forest &	\textbf{64.51}\% &	\textbf{67.74}\% &	\textbf{72.04}\% &	\textbf{0.588} &	\textbf{0.676} \\ 
         & 	DNN &	\textbf{68.81}\% &	\textbf{73.11}\% &	\textbf{75.26}\% &	\textbf{0.612} &	\textbf{0.708} \\ 
         & 	ENET &	9.67\% &	46.23\% &	58.06\% &	0.2212 &	0.2537 \\ 
         &	GBoost &	\textbf{62.36}\% &	\textbf{73.11}\% &	\textbf{77.41}\% &	\textbf{0.592} &	\textbf{0.677} \\ 
         &	XGBoost &	63.44\% &	70.96\% &	77.41\% &	0.597 &	0.678 \\ 
         &	Lasso &	9.67\% &	45.16\% &	58.06\% &	0.222 &	0.255 \\ 
         &	LGBM &	64.51\% &	70.96\% &	74.19\% &	0.592 &	0.68 \\ \hline
        SMOTE oversampling &	Random Forest &	\textbf{61.29}\% &	\textbf{67.74}\% &	\textbf{69.89}\% &	\textbf{0.553} &	\textbf{0.645} \\ 
         & 	DNN &	\textbf{39.78}\% &	\textbf{58.06}\% &	\textbf{66.66}\% &	\textbf{0.439} &	\textbf{0.497} \\ 
         & 	ENET &	18.27\% &	44.08\% &	55.91\% &	0.275 &	0.308 \\ 
         &	GBoost &	\textbf{60.21}\% &	\textbf{75.26}\% &	\textbf{76.34}\% &	\textbf{0.587} &	\textbf{0.662} \\ 
         &	XGBoost &	60.21\% &	68.81\% &	73.11\% &	0.563 &	0.651 \\ 
         &	Lasso &	18.27\% &	44.08\% &	55.91\% &	0.275 &	0.308 \\ 
         &	LGBM &	60.21\% &	73.11\% &	74.19\% &	0.569 &	0.657 \\ \hline
    \end{tabular}
    
\end{table*}

\begin{table*}[!htbp]
    \caption{Performance comparison of Combined model of classifier and regressor with Regressor on Tomcat project}
    \label{tab:exp3}
    \centering
    \begin{tabular}{|c|p{0.25\linewidth}|c|c|c|c|c|}
    \hline
        \textbf{\textit{Model Name}}  & \textbf{\textit{ Oversampling technique used}} & \textbf{\textit{accuracy@1}} & \textbf{\textit{accuracy@5}} & \textbf{\textit{accuracy@10}} & \textbf{\textit{MAP}} & \textbf{\textit{MRR}} \\ \hline
        Random Forest &	No Oversampling &	\textbf{64.51\%} &	\textbf{67.74\%} &	\textbf{72.04\%} &	\textbf{0.588} &	\textbf{0.676} \\ \hline
        Random Forest &	Classifier and SMOTE oversampling &	62.17\% &	67.91\% &	69.89\% &	0.5712 &	0.663 \\ \hline
        Random Forest &	SMOTE oversampling &	61.29\% &	67.74\% &	69.89\% &	0.553 &	0.645 \\ \hline
    \end{tabular}

\end{table*}

We train our dataset using each of these models by using SMOTE as well as without Oversampling to verify which model gives us the best results for our task of Bug Localization. Table \ref{tab:exp2} shows the results of all the above models with and without oversampling. From this, we can observe that the DNN regressor, Gradient boost, and Random Forest regressor yields better results for our task without oversampling. So, we perform our next set of experiments only on these models.

\subsection{Combined Model of Classifier and Regressor}    
To remove class imbalance from the training data, we experimented with various over-sampling as well as under-sampling techniques. Among all other techniques, SMOTE gave us the best results. \par

There can be some data elements of positive pairs in our training data, which are wrongly interpreted as not buggy after training the random forest model. However, while oversampling, we are considering those wrongly interpreted elements also to remove class imbalance, which might spoil our results. So we built a random forest classifier from the same training dataset, which classifies each data element as buggy or not buggy. We remove all the wrongly classified data elements from our dataset and then apply SMOTE for oversampling. Then random forest regressor model is trained on this data and evaluated on test data. Table \ref{tab:exp3} shows a performance comparison between random forest model without oversampling, with oversampling by SMOTE technique and combined model of random forest classifier and regressor with oversampling by SMOTE technique.

\begin{figure}
\centering
\includegraphics[width=\linewidth]{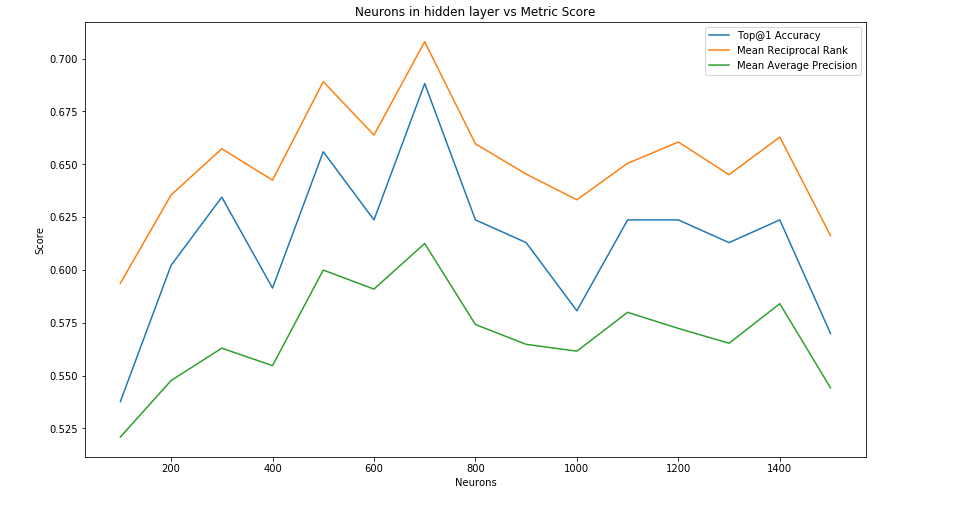}
\caption{accuracy@1, MAP, MRR vs Number of Neurons}
\label{fig:exp4}
\end{figure}

\subsection{Deep Neural Network Model}
Performance of Deep Neural Network (DNN) regressor changes by varying the number of hidden layers along with the number of neurons present in it, as shown in Figure \ref{fig:exp4}. As we have only six features to train our DNN model, we decided to add only one hidden layer in our model. So our DNN model has 3 layers in total, which are the input layer with six neurons for six features, a hidden layer, and an output layer with one neuron for a single output. 

We perform an experiment to find a number of hidden layer neurons to achieve the best results. We vary the number of neurons present in the hidden layer from 100 to 1500 by increasing at a rate of 100 at each step, evaluating each model, and plotting the graph between MAP, MRR, accuracy@1, and the number of neurons in the hidden layer. Figure \ref{fig:exp4} represents that plot, and we can observe that we get the best results when the number of neurons in the hidden layer is 700. So we decided to use 700 neurons in the hidden layer for further experiments.

\begin{table*}[!htbp]
    \caption{Results of DRAST on C projects in BugC}
    \label{tab:result1}
    \centering
    \begin{tabular}{|c|c|c|c|c|c|c|}
    \hline
        \textbf{\textit{Project}} & \textbf{\textit{Model}} & \textbf{\textit{accuracy@1}} & \textbf{\textit{accuracy@5}} & \textbf{\textit{accuracy@10}} & \textbf{\textit{MAP}} & \textbf{\textit{MRR}} \\ \hline
        
        Systemd &	DRAST - RandomForest &	92.01\% &	97.17\% &	97.74\% &	0.952 &	0.973 \\ 
          &	DRAST - DNN &	90.31\% &	96.69\% &	97.74\% &	0.924 &	0.939 \\ 
          &	DRAST - GBoost &	92.65\% &	97.74\% &	98.30\% &	0.943 &	0.959 \\ \hline
        rsyslog  &	DRAST - RandomForest &	95.55\% &	96.66\% &	96.66\% &	0.952 &	0.959 \\ 
          &	DRAST - DNN &	95.55\% &	95.55\% &	96.66\% &	0.936 &	0.945 \\ 
          &	DRAST - GBoost &	94.44\% &	95.55\% &	97.77\% &	0.947 &	0.953 \\ \hline
        radare2  &	DRAST - RandomForest &	91.01\% &	96.62\% &	96.62\% &	0.9143 &	0.9199 \\ 
          &	DRAST - DNN &	88.76\% &	95.50\% &	95.50\% &	0.9064 &	0.913 \\ 
          &	DRAST - GBoost &	92.13\% &	97.75\% &	97.75\% &	0.9373 &	0.9432 \\ \hline
        lightning  &	DRAST - RandomForest &	94.44\% &	100\% &	100\% &	0.941 &	0.959 \\ 
          &	DRAST - DNN &	92.59\% &	100\% &	100\% &	0.923 &	0.957 \\ 
          &	DRAST - GBoost &	94.44\% &	100\% &	100\% &	0.941 &	0.959 \\ \hline
         
          neomutt & DRAST - RandomForest & 94.74\% & 100.00\% & 100.00\% & 0.977	& 0.974 \\
          & DRAST - DNN & 97.37\% &	100\% &	100\% &	0.961 &	0.949 \\
          & DRAST - GBoost & 94.74\% &	100.00\% &	100.00\% &	0.977 &	0.974 \\ \hline
          
          pygame & DRAST - RandomForest & 96.43\% &	96.43\% &	100.00\% &	0.775 &	0.969 \\
          & DRAST - DNN & 100\%	& 100\%	& 100\%	& 0.795 & 1 \\
          & DRAST - GBoost & 100.00\% & 100.00\% &	100.00\% &	0.8 &	1 \\ \hline
          
          citus & DRAST - RandomForest & 96.82\% & 96.82\%	& 96.82\%	& 0.943 & 0.971 \\
          & DRAST - DNN & 93.65\% &	95.24\%	& 96.82\% &	0.911 &	0.948 \\
          & DRAST - GBoost & 95.24\%	& 96.82\% &	100.00\%	& 0.929 & 	0.964 \\

          \hline
    \end{tabular}

\end{table*}

\begin{table*}[!htbp]
    \caption{Comparision of DRAST with state-of-the-art techniques}
    \label{tab:benchmark}
    \centering
    \begin{tabular}{|c|c|c|c|c|c|c|}
    \hline
        \textbf{\textit{Project}} & \textbf{\textit{Model}} & \textbf{\textit{accuracy@1}} & \textbf{\textit{accuracy@5}} & \textbf{\textit{accuracy@10}} & \textbf{\textit{MAP}} & \textbf{\textit{MRR}} \\ \hline
        Tomcat &	DRAST &	68.81\% &	73.11\%	& 75.26\% &	0.612 &	0.708 \\ 
         &	BugLocator &	35.40\% &	64.50\% &	70.90\% &	0.431 &	0.485 \\ 
         &	DNNLOC &	53.90\% &	72.90\% &	80.40\% &	0.52 &	0.6 \\ 
         &	DeepLocator	& 52\% &	72\% &	80\% &	0.54 &	0.6 \\ 
         &	NPCNN &	53\% &	70\% &	79.20\% &	0.529 &	0.597 \\ 
         &	CAST &	50.70\% &	76.60\% &	82.50\% &	0.556 &	0.612 \\ \hline
         AspectJ &	DRAST &	70.00\% &	70.00\% &	70.00\% &	0.662 &	0.700 \\
         &	BugLocator &	21.60\% &	47.30\% &	57.10\% &	0.276 &	0.369 \\
         &	DNNLOC &	47.80\% &	71.20\% &	80.40\% &	0.32 &	0.52 \\
         &	DeepLocator &	40.00\% &	66.00\% &	78.00\% &	0.34 &	0.49 \\
         &	NPCNN &	46.00\% &	73.00\% &	81.00\% &	0.401 &	0.531 \\
         &	CAST &	50.00\% &	76.60\% &	83.00\% &	0.418 &	0.536 \\ \hline
    \end{tabular}
    
\end{table*}

\section{Results and Discussion} \label{section 8}
In this section, we discuss the performance of the DRAST on various projects in BugC and benchmark dataset and compare our results with the state-of-the-art techniques. All the results discussed in this section are performed \textit{without oversampling} as, comparatively, the results with oversampling are not better (Section \ref{section7}). We tried different models and different oversampling techniques. However, due to limited resources and time, we are not able to perform different experiments by varying the parameters like the number of nodes for finding/improving the DNN relevancy score, which could be performed as part of future work. It could also be one of the reasons for the slow increase in accuracy@k when compared to the state-of-the-art techniques (see Table \ref{tab:benchmark}). 

\subsection{Performance of DRAST on BugC Dataset}
We applied DRAST on seven projects from the BugC dataset and two projects from the benchmark dataset by varying the final model with RandomForest, DNN, and GBoost. Table \ref{tab:benchmark} shows the results for Tomcat and AspectJ (projects from Java benchmark dataset), and Table \ref{tab:result1} shows the results for seven C projects from the BugC dataset. We can observe that the results of DRAST on C projects are better when compared to the Java projects. Accuracy@k for C projects is at least 90\%. i.e., out of thousands of source code files present in the project, the developer only needs to check the first file using DRAST to fix the bug in 90\% of the cases. We observe that the DRAST model in which DNN is used for the final regression task works better for Java project tomcat, and for the C projects, GBoost gave good results. 

\subsection{Comparison with State-of-the-Art Techniques}
DRAST is compared on the Tomcat and AspectJ projects from the benchmark dataset with five state-of-the-art bug localization techniques, namely BugLocator, DNNLOC, DeepLocator, NPCNN, and CAST. Table \ref{tab:benchmark} shows the results of different metrics on the above-mentioned techniques and DRAST with the DNN model. Results show that DRAST outperforms all five bug localization at accuracy@1, MAP, and MRR by a huge margin. 

The huge gap in accuracy@1 values between DRAST and other state-of-the-art techniques could be due to the comparison of source code with bug reports at the code blocks level, whereas other techniques are comparing at the file level. The similarity of bug report at the file level may be less for some files due to their large size, but at the code block level, the similarity might be high, and the chances of finding buggy code blocks can be more in such cases.

DRAST is better than BugLocator in accuracy@1,5,10, MAP, and MRR, because it tries to reduce the lexical gap between source code features and bug reports using the DNN relevancy score. DRAST performance at accuracy@1,5 is better than accuracy@10 with respect to all the state-of-the-art bug localization techniques. We observed that it requires a huge amount of time and memory to conduct a single experiment, and changing hyper-parameters might reveal better results for accuracy@10 metric. Subject to these constraints, we tried to fix the parameters according to those mentioned in the literature \cite{lam2017bug}.


\section{Threats to Validity}
\begin{itemize}
    \item \textbf{Dataset}: We tried DRAST on the current benchmark dataset and on 7 out of 21 projects from BugC dataset. Performance of DRAST is dependent on the curation process of the dataset, which includes design decisions such as choosing open source GitHub projects, considering issues as bugs, and relying on GitHub API in linking the issues and pull requests.
    \item \textbf{Word Embedding:} Out of many available word embedding techniques, we chose to represent the bug reports and code features as tf-idf vectors. There might be other word embedding techniques that might work well for our model. We explore those techniques for future studies. 
    \item \textbf{Dimensionality Reduction}: The choice of reducing the dimensions to 75\% might have an effect on the final results. 
    \item \textbf{Use of rVSM}: As mentioned in the literature, we used the rVSM model to calculate the textual similarity. Therefore the final output score depends on the rVSM model. 
    \item \textbf{srcML}: For extracting code blocks, which play a major role in our model, we depended on the output of srcML and its documentation, and also, there is a chance that we might have missed a few details while converting the XML to code blocks.
    \item \textbf{Textual similarity cutoff:} To reduce the number of pairs of bug reports and files, we only considered those pairs whose textual similarity feature is greater than the cutoff (0.1). Final results may also be affected by the value of the cutoff we choose, and further studies will explore the optimal cutoff to improve the performance of DRAST. 
    \item \textbf{Fold wise metrics}:  Dataset is dividing into folds to reduce the time for training model, results we get while testing on different folds are varying, and we have reported maximum we get while testing. This variance is not considered while evaluating the model. 
    \item \textbf{Choice of Projects}: The selected seven C projects are choosen randomly from BugC dataset. Also, the DRAST often takes a lot of time to achieve the results, so we limit ourselves to seven projects because most state-of-the-art methods have only been evaluated on 5-6 projects \cite{zhou2012should, liang2019deep, xiao2017improving, lam2017bug, huo2017enhancing}. The results that we got are dependent on the quality of the project and the bug reports. 
\end{itemize}

\section{Related Work}
Bug localization approaches are majorly divided into two fundamental approaches: Information Retrieval (IR) and Deep Learning Techniques:
\subsection{IR Based Approach}
Information retrieval is a technique used for extracting important and relevant information from resources. PROMESIR is developed based on Latent Semantic Indexing (LSI) and probabilistic ranking method, locating bugs using cosine similarity obtained from source code and bug reports \cite{poshyvanyk2007feature, poshyvanyk2006combining}. Lukins et al. \cite{lukins2008source} used Latent Dirichlet Allocation (LDA), a topic modeling model where topic distributions extracted from bug reports and source code is compared, while in BugScout \cite{nguyen2011topic} an extended LDA model introduce defect-proneness factor which considers frequently fixed files. Vector space model (VSM) is used to compute the similarity between the source code files and bug reports \cite{gay2009use}. On similar lines, another study is conducted to compare IR-based techniques on bug localization on the iBugs dataset and found that VSM performed better than LDA, LSA (latent semantic analysis), and CBDM (cluster-based document model) \cite{rao2011retrieval}. Zhou et al. \cite{zhou2012should} proposed a model called BugLocator, based on a revised vector space model (rVSM) that considers the historical reports of similar bugs that have been fixed before and their corresponding buggy files to improve the ranking performance. BRTracer \cite{wong2014boosting}, a bug localization tool and an extension of BugLocator, uses those segments of the source code (segmentation) to represent the source code file, which is more similar to the bug report. It also gave importance to the files that are available in bug reports as stack traces (stack trace analysis) \cite{wong2014boosting}. Segmentation and stack trace analysis are complementary to each other. Although it is convenient to correlate the diversified lexical data in the same feature space, the approaches mentioned above usually require a lot of computational capability and mainly dependent on the features extracted from the source code and bug reports \cite{lecun2015deep}. 

BLUiR tool considers class and method names and similar bug data for bug localization \cite{saha2013improving, saha2014effectiveness}. Ye et al. \cite{ye2014learning} introduced the learning-to-rank approach in which given a bug report, features are extracted from source code, API descriptions, code change, and bug-fixing history are used to compute weights for each file to list out probable buggy files. 
A large-scale comparison between IR-based on Bug Localization is made to identify the performance of newer tools with the older ones \cite{akbar2020large}. The authors divide the IR-based tool into three generations. They are evaluated against 20,000 bug reports taken from Java, C, C++, and Python projects, and the results found that third-generation IR based tools and the use of word embeddings in these techniques are effective for bug localization purposes \cite{akbar2020large}. Recently, Laprob \cite{li2020laprob} a graph-based method that uses the label-propagation method by constructing a Biparty Hybrid Graph that analyzes inter and intra-relations between bug reports and source code. Laprob is tested against several IR-bases bug localization techniques and showed promising results.

Some studies also focus on the query aspect of bug reports in the context of bug localization \cite{mills2020relationship,zhang2019finelocator}. Chris et al. \cite{mills2020relationship} argue that better query can be formulated using the bug report, which can help achieve better results for IR based approaches. FineLocator performed method-level bug localization by considering three query expansion scores to address the "representation sparseness problem" of a function in the source code \cite{zhang2019finelocator}. One of the main limitations of these IR based techniques is the lexical mismatch between bug reports (written in natural language) and source code  \cite{xiao2017improving, lam2017bug, huo2019deep}.

\subsection{Deep Learning-Based Approaches}
Recent work in bug localization in which machine learning or deep learning (DL) techniques reduce the lexical and semantic gap between a bug report and source code by representing them in high-dimensional features. This is evident from the results where DL techniques outperform IR based bug localization approaches. 
A two-phase model is proposed where bug reports are classified as ``predictable'' or ``deficient'' using Naive Bayes, and prediction is made on the former type \cite{kim2013should}. Naive Bayes use features extracted from bug reports consist of metadata such as version, platform, and previously fixed file history for locating the file. Lam et al. developed HyLoc \cite{lam2015combining} and DNNLOC \cite{lam2017bug} models, which achieves greater accuracy by combining the features extracted from DNN, rVSM, and the project's bug-fixing history. However, the techniques mentioned above do not take structural information from source code into consideration. Studies conducted by Huo et al.  \cite{huo2016learning} and \cite{huo2017enhancing} proposed NP-CNN, which extracts structural information of the source code using Convolution Neural Network (CNN) and learn unified features from source code and bug reports. It also solves the problem of lexical mismatch between source code written in programming language and bug reports written in natural language. Deeplocator  \cite{xiao2017improving} achieves a 3.8\% higher MAP than DNNLOC using an enhanced CNN rather than considers bug-fixing history.  Xiao et al. \cite{xiao2019improving} proposed DeepLoc, which represents the source code and bug reports in a word embedding form and used enhanced CNNs to detect the features between them. Knowledge Graphs are also leveraged in bug localization, such as KGBugLocator \cite{zhang2020exploiting}, which uses knowledge graph embeddings and Bi-Directional Attention Mechanism to extract information between bug reports and source code.

Liang et al. \cite{liang2019deep} proposed a bug localization method called CAST that exploits deep learning and customized AST's of source code to automatically and effectively locate potential buggy source files. In particular, CAST extracts lexical semantics from both source files (e.g., method names) and bug reports (e.g., words). As they are using customized AST by compressing AST's size, there is a chance to lose some important information for locating bugs during that compression. TRANP-CNN, one of the first approaches in cross-project bug localization, uses CNNs for transfer learning to extract ``transferable features'' from code files of source and target projects and source project bug reports for bug localization \cite{huo2019deep}. CooBA \cite{DBLP:conf/ijcai/Zhu0T020} an enhancement of TRANP-CNN is a cross-project bug localization approach which used adversarial transfer learning to avoid negative transfer between projects and provide an effective extraction of shared information between multiple projects.

\section{Conclusion and Future Work}

The goal of Bug Localization is to help developers by making their task of finding buggy files relatively easy and with less effort in the context of large scale projects. We developed \textit{\textbf{DRAST}}, a bug localization approach for projects developed in \textit{C} and \textit{Java} language and \textit{\textbf{BugC}}, a C language dataset to test our approach. DRAST is developed by using \textit{src2vec}, a high-level AST representation for source code, and combining Information Retrieval technique (rVSM) with Machine Learning and Deep Learning models such as Random Forest, Gradient boost, and Deep Neural Network regressor. It outputs the ranked list of probable buggy files in descending order.

BugC consists of 2462 bug reports drawn from open-source GitHub projects written in C programming language and consists of information that includes bug reports and pull-requests. \textit{src2vec} is a novel approach for representing Java and C language source file as code vectors by using high level AST using srcML.

DRAST is tested on seven C projects from BugC dataset, and two Java projects, i.e., Tomcat and AspectJ from benchmark dataset to compare the DRAST's performance with state of the art models, and our results suggest that DRAST significantly outperforms other models in terms of \textit{accuracy@1}, \textit{MAP} and \textit{MRR} metrics. Results shows that DRAST works better on C projects compared to Java projects. The results obtained states that in nearly 90\% of cases in C and 60\% of cases in Java, the developer will need to check just one file out of thousands of files available in the projects. 

DRAST, as of now, supports C and Java projects, but this can be extended to work even for projects developed in other programming languages such as C++ and Python, by creating XML parser for that language by following srcML documentation. Many experiments can be performed to decide DNN regressor hyper-parameters and the number of folds required in the training data. DRAST currently can perform file-level bug localization and it can be extended for method level bug localization tasks as well.

\bibliographystyle{elsarticle-num}
\bibliography{main}
\end{document}